\documentclass[twocolumn,usenatbib]{mn2e}
\usepackage{graphicx,natbib}
\usepackage{epsfig}

\title[Emergences of C IV narrow absorption troughs]{Emergences of C IV narrow absorption troughs in the quasar
SDSS J095254.10+021932.8}
\author[Zhi-Fu Chen et al.]
{Zhi-Fu Chen$^{1, 2}$\thanks{E-mail:zhichenfu@126.com}, Mu-Sheng Li$^2$, Wei-Rong Huang$^2$, Cai-Juan Pan$^1$, You-Bing Li$^3$\\
$^{1}$ Department of Physics and Telecommunication Engineering, Baise University, Baise, 533000, China\\
$^{2}$ Center for Astrophysics, Guangzhou University, Guangzhou 510006, China\\
$^{3}$ City Construction College of Guangzhou, Guangdong, 510925,
China\\}
\begin{document}
\maketitle

\begin{abstract}
In this paper, we report on two C IV narrow absorption systems
emerging from the spectrum of the quasar SDSS J095254.10 +021932.8,
which are located at $z_{\rm abs}=2.0053$ and $z_{\rm abs}=1.8011$,
respectively. These features have velocity offsets of  $\rm \sim
13,800$ and $\rm \sim 34,700~km~s^{-1}$ with respect to the quasar,
and show obviously partial coverage to the background emission
sources. The two C IV absorption systems are imprinted on the
spectrum obtained by the Sloan Digital Sky Survey (SDSS) III on 2011
March 25, which cannot be observed from the spectrum obtained by
SDSS-I/II on 2000 December 30. The time interval of the two SDSS
observations is 1186.4 d at the quasar rest frame. Because the
continuum radiation, broad emission lines and mass accretion rate
are stable for the two observations, together with the time interval
of observations, we believe that the observed spectral variations
are less likely to be caused by the changes in the ionization states
of the absorbing gas or by a new outflow, arising from the situation
where there was previously no outflow at all. Instead, they probably
arise from the motions of multiple streaming gas across the
sightline of the quasar.
\end{abstract}

\begin{keywords}
galaxies: active --- quasars: absorption lines --- quasars:
individual (J095254.10+021932.8)
\end{keywords}

\section{Introduction}
Active galaxies are powered by accreting material on to the
corresponding central supermassive black holes. The gas not only
falls into the black hole, but can also be lifted off the accretion
disc under some mechanisms in powerful winds/outflows. Outflows seem
to be the fundamental component of quasar systems, and are present
in at least 50 per cent of optically selected quasars (Crenshaw,
Kraemer \& George 2003; Ganguly \& Brotherton 2008; Nestor, Hamann
\& Rodr\'Iguez Hidalgo 2008). It is widely accepted that the
outflows play an important role in regulating the growth of
supermassive black holes by carrying away the angular momentum (Silk
\& Rees 1998; King 2003). Moreover, the outflows might inject enough
kine-matic energy and metal-rich gas into the host galaxies,
influencing the star formation, the enrichment of the surrounding
intergalactic medium, and so on (e.g. Di Matteo, Springel \&
Hernquist 2005; Bower et al. 2006; Moll et al. 2007; Dunn et al.
2012). Therefore, it is important to understand the properties and
evolution of quasar outflows.

Unfortunately, the nature and origin of quasar outflows are still
mysterious. Some models suggest that the quasar outflows are
physically related to the gas lifted off the accretion disc, which
would be driven by the radiation pressure, magnetic fores, or a
combination of these (Arav, Li \& Begelman 1994; de Kool \& Begelman
1995; Murray et al. 1995; Murray \& Chiang 1997; Proga, Stone \&
Kallman 2000; Everett 2005). Here, the radiation pressure seems to
play a dominant role (Laor \& Brandt 2002; Hamann et al. 2011). The
quasar outflows are detected most conspicuously via blueshifted
metal absorption lines. These absorption lines imprint on the quasar
spectra in different forms based on linewidths: broad absorption
lines (BALs) with absorption troughs that are broader than
$2000~km~s{-1}$ at depths $>10$ per cent below the continuum; narrow
absorption lines (NALs) with velocity widths of a few hundred $\rm
km~s^{-1}$; intermediate width absorption lines (mini-BALs; e.g.
Weymann et al. 1991; Jannuzi et al. 1996; Yuan et al. 2002; Wise et
al. 2004; Misawa et al. 2008; Hamann et al. 2011). It is possible
that the different types of lines arise in the same quasar outflow
but are viewed at different angles (Murray et al. 1995).
Alternatively, they might be formed in the same quasar outflow but
at different evolutionary stages, where NALs and mini-BALs appear
near the beginning or end stages of a more powerful BAL outflow
(e.g. Hamann \& Sabra 2004; Hamann et al. 2008).

BALs are undoubtedly intrinsic to quasars. Therefore, many pre-vious
studies of outflow absorption lines have focused on BALs (e.g.
Gibson et al. 2008; Moe et al. 2009; Capellupo et al. 2011, 2012,
2013). However, it is difficult to distinguish intrinsic NALs from
cosmologically intervening NALs, which are often believed to be
associated with cosmologically intervening galaxies. In spite of
this difficulty, many works on quasar outflows have been carried out
using NALs (e.g. Narayanan et al. 2004; Misawa et al. 2007; Hamann
et al. 2011; Chen, Qin \& Gu 2013). These intrinsic absorption
lines, which are highly ionized (e.g. O VIII,Fe XVII), are mainly
detected in the high-energy band spectra (e.g. X-ray spectra), while
the transitions by ions with low ionization potential (e.g. C IV,S
IV, N V ,O VI) can be observed in both the X-ray and ultraviolet
(UV) spectra. The absorption lines, which are imprinted in both the
UV and X-ray spectra with similar velocities relative to the
quasars, imply an underlying relationship between the narrow
absorption lines in the UV and X-ray spectra (Mathur et al. 1994;
Mathur, Elvis \& Wilkes 1995; Sabra et al. 2003; Tombesi et al.
2011a,b; Gupta et al. 2013a).

Line variability is likely to be a good method for investigating
quasar outflows, because it can constrain the outflow dynamics,
stability, location and basic physical properties. The variability
of absorption lines imprinted on the quasar spectra could be related
to the fluctuation of the background emission sources. This would
induce a variation in the ionization states of the absorbing gas
and/or the proper motion of absorbers, which could give rise to
changes in the column density or changes in covering factors.

Time variations of BALs and NALs are relatively common, but the
extreme changes in the absorption profiles in the near-UV and
optical spectra are rare. We are aware that only a few cases of the
disappearance (e.g. Junkkarinen et al. 2001; Lundgren et al. 2007;
Hall et al. 2011; Filiz Ak et al. 2012) and emergence (Ma 2002;
Hamann et al. 2008; Leighly et al. 2009; Krongold, Binette \&
Hernandez-Ibarra 2010; Rodr\'Iguez Hidalgo, Hamann \& Hall 2011;
Vivek et al. 2012) of BALs have been reported in the near-UV and
optical spectra. These can be detected on time-scales of years at
the quasar rest frame.

It would be very interesting to detect the emergence of absorption
lines because the variability programmes usually monitor quasars for
which the absorption lines have already been detected. In this
paper, we report on two emergence events of CIV absorption sys-tems,
which are imprinted on the spectrum of the quasar SDSS J095254.10
+021932.8. In Section 2, we provide a spectral analy-sis, and we
present a discussions and conclusions in Sections 3 and 4,
respectively. Throughout this paper, we adopt the cosmological
parameters $\Omega_\Lambda=0.7$, $\Omega_M=0.3$, and $h=0.7$.

\section{Spectral analysis}
The Baryon Oscillation Spectroscopic Survey (BOSS), which is part of
the Sloan Digital Sky Survey (SDSS) III (Eisenstein et al. 2011),
aims to obtain the spectra of over 150 000 quasarsz $z>2.2$ (Ross et
al. 2012). The SDSS Data Release 9 is the first release of BOSS
spec-troscopy to the public, which contains $\sim 87,000$ quasars
detected over 3,275 $deg^2$ (P\^aris et al. 2012). The 7932 quasars
obtained by SDSS-I/II (York et al. 2000; Abazajian et al. 2009;
Schneider et al. 2010) have been reobserved during the two first
years of BOSS.

Quasar spectra with a higher signal-to-noise ratio (S/N) are useful
for the detection of weak absorption lines. We aim to detect
$C~IV~\lambda\lambda1548,1551$  absorption doublets on quasar
spectra with high S/N and with two observations. Specifically, we
select quasars from the SDSS quasar catalogue, which have been
observed by both SDSS-I/II and SDSS-III, and whose median S/N over
the whole spectrum is greater than 15 (the median S/N over the whole
spec-trum is defined by P\^aris et al. 2012). Because of the
significant systematic sky-subtraction residual longward of 7000
\AA~ and the noise region shortward of 4000 \AA~ in many SDSS-I/II
spectra, we constrain the wavelength range detected
$C~IV~\lambda\lambda1548,1551$  absorption doublets from $4,000$
\AA~ to $7,000$ \AA~ at observed frame. The $\rm O~I~\lambda1302$
and $\rm Si~II~\lambda1304$ pair has a wavelength separation similar
enough to that of the $\rm C~IV~\lambda\lambda1548,1551$. In order
to avoid the the $\rm Ly \alpha$ forest, $\rm O~I~\lambda1302$ and
$\rm Si~II~\lambda1304$ absorption lines, we exclude the spectral
region shortward of 1350 \AA~ at rest-frame. Based on above
selection criteria, there are 680 appropriate quasars to search for
$\rm C~IV~\lambda\lambda1548,1551$ absorption doublets.

Following some previous works (e.g. Nestor, Turnshek \& Rao 2005;
Quider et al. 2011; Chen et al. 2013), we invoke the cubic splines
(for the underlying continuum, see Willian et al. 1992 for details)
plus the Gaussians (for the emission and broad absorption figures)
to fit a pseudo-continuum for each of the 680 quasar spectra. This
is used to normalize the spectral fluxes and flux uncertainties. The
process is iterated several times to improve the fits of both the
cubic splines and the Gaussians. We detect absorption figures on the
pseudo-continuum normalized spectra. During the detecting, we first
mask out the three times flux uncertainty levels ($3\sigma$), which
have been normalized by the pseudo-continua, and we rule out the
absorption figures located within $3\sigma$. In the sec-ond step, we
flag the absorption figures imprinted on the SDSS-III spectra but
not on those of SDSS-I/II, and we rule out absorption figures with
absorption troughs that are broader than $1000~{\rm km~s^{-1}}$ at
depths $>10\%$ below the continuum. In the third step, we invoke a
Gaussian component to fit each absorption trough, and we rule out
the trough with $FWHM>600~{\rm km~s^{-1}}$. Then, in the fourth
step, we search for the $C~IV~\lambda\lambda1548,1551$. We find only
one doublet, which was imprinted on the quasar spectrum of
J095254.10 +021932.8. This $C~IV~\lambda\lambda1548,1551$ absorption
doublet (the $R-component$ shown in Fig. 2) is located at $z_{\rm
abs}=2.0053$ and has $FWHM_{\rm \lambda1548}=420~{\rm km~s^{-1}}$.
The spectra of this quasar are plotted in Fig. 1, where the red
solid lines represent the pseudo-continua fittings. In Fig. 2, we
show the pseudo-continuum normalized spectra. For this C IV
absorption system, we expect to detect the absorptions caused by
other metal species with relatively high abundances of elements.
However, we find that there are only two appropriate lines ($\rm
S~I~\lambda1444$ and $\rm Co~II~\lambda1448$), which have fallen
into both the SDSS-I/II and SDSS-III spectra. However, many
appropriate lines are located at the $\rm Ly\alpha$ forest region
and their wavelengths are not covered by the SDSS-I/II spectra. We
invoke the Gaussian function to measure the equivalent widths at the
rest frame of the detected absorption lines and we estimate their
uncertainties using the Gaussian fittings. The fitting results are
shown in Fig. 2 and presented in Table 1.

The question is whether the whole trough near to 1380 ? at the
quasar rest frame, which was imprinted on the SDSS-III spectrum but
not on that of the SDSS-I/II, was mainly caused by the absorp-tion
of the $\rm Si~IV~\lambda\lambda1393,1402$ doublet. When the $\rm
Si~IV~\lambda\lambda1393,1402$ absorption doublet is located at the
spectra region near to $1380$ \AA~ at quasar rest-frame,  the
wavelength separation of the $Si~IV~\lambda\lambda1393,1402$ doublet
is $\sim9$ \AA.~ However, the width of this whole absorption trough
is only $7.3$ \AA~ at depths $>10\%$ below the continuum. Therefore,
this trough is unlikely to be mainly caused by the absorption of the
$\rm Si~IV~\lambda\lambda1393,1402$ doublet. It can clearly be seen
from Fig. 2 that the absorptions of $\rm S~I~\lambda1444$ and $\rm
Co~II~\lambda1448$ alone, which are consistent with the R-component,
cannot completely account for this trough. We invoke four Gaussian
components to fit this trough, of which two account for the
absorptions of $\rm S~I~\lambda1444$ and $\rm Co~II~\lambda1448$,
which are consistent with the R-component. We find that the other
two Gaussian components are consistent with the absorption of the
$\rm C~IV~\lambda\lambda1548,1551$ doublet located at $z_{\rm
abs}=1.8011$ . The fitting results are shown in Fig. 2 and presented
in Table 1.

We also measure other obvious absorption lines, which are imprinted
on both the SDSS-I/II and SDSS-III spectra of this quasar. These
absorption lines can be divided into five intervening $\rm
Mg~II~\lambda\lambda2796,2803$ absorption systems. The fitting
results are plotted in Fig. 2 and presented in Table 2.

\begin{table}
\centering\caption{Measurements of C IV absorption systems} \tabcolsep 2mm 
 \begin{tabular}{cccccc}
 \hline\hline\noalign{\smallskip}
Species&$z_{\rm abs}$&$W_r$ [\AA]&$W_r^{\rm a}$ [\AA]\\
 \hline
$C~IV \lambda1548$&2.0053&$0.70\pm0.12$&0.13\\
$C~IV \lambda1551$&&$0.60\pm0.10$&0.11\\
$S~I \lambda1444$&&$0.29\pm0.05$&0.07\\
$Co~II \lambda1448$&&$0.42\pm0.07$&0.10\\
$N~V \lambda1238$&&$0.95\pm0.13$&$\cdots$\\
$N~V \lambda1242$&&$1.51\pm0.19$&$\cdots$\\
$Si~II \lambda1260$&&$1.24\pm0.08$&$\cdots$\\
$Si~I \lambda1255$&&$1.11\pm0.17$&$\cdots$\\
$S~II \lambda1253$&&$0.58\pm0.12$&$\cdots$\\
$S~II \lambda1250$&&$0.36\pm0.09$&$\cdots$\\
\hline
$C~IV \lambda1548$&1.8011&$0.24\pm0.08$&0.08\\
$C~IV \lambda1551$&&$0.33\pm0.08$&0.08\\
\hline\hline\\
\end{tabular}
\\
$^a$The equivalent width limits for the SDSS-I/II spectra.
\end{table}

\begin{table}
\centering\caption{Measurements of Mg II absorption systems} \tabcolsep 2mm 
 \begin{tabular}{cccccc}
 \hline\hline\noalign{\smallskip}
Species&$z_{abs}$&$W_r^a$ [\AA]&$W_r^b$ [\AA]\\
 \hline
$Mg~II\lambda2796$&0.3635&$1.25\pm0.15$&$0.91\pm0.20$\\
$Mg~II\lambda2803$&&$1.18\pm0.15$&$0.81\pm0.25$\\
\hline
$Mg~II\lambda2796$&0.4167&$0.82\pm0.13$&$0.48\pm0.13$\\
$Mg~II\lambda2803$&&$0.64\pm0.12$&$0.78\pm0.17$\\
\hline
$Mg~II\lambda2796$&0.9634&$0.92\pm0.07$&$1.17\pm0.16$\\
$Mg~II\lambda2803$&&$0.82\pm0.08$&$0.77\pm0.10$\\
$Fe~II\lambda2600$&&$0.53\pm0.07$&$0.52\pm0.12$\\
$Fe~II\lambda2587$&&$0.16\pm0.05$&$0.24\pm0.10$\\
$Fe~II\lambda2383$&&$0.49\pm0.07$&$0.56\pm0.14$\\
$Fe~II\lambda2344$&&$0.25\pm0.06$&$0.25\pm0.12$\\
$Mn~I\lambda2004$&&$0.47\pm0.14$&$0.18\pm0.07$\\
\hline
$Mg~II\lambda2796$&1.1484&$0.39\pm0.08$&$0.27\pm0.10$\\
$Mg~II\lambda2803$&&$0.23\pm0.05$&$0.19\pm$0.06\\
\hline
$Mg~II\lambda2796$&1.3352&$0.69\pm0.06$&$0.78\pm0.10$\\
$Mg~II\lambda2803$&&$0.49\pm0.07$&$0.63\pm0.10$\\
$Al~III\lambda1854$&&$0.26\pm0.05$&$0.30\pm0.07$\\
$Al~ II\lambda1670$&&$0.14\pm0.04$&$0.12\pm0.07$\\
\hline\hline\\
\end{tabular}
\\
$^a$The measurements from the SDSS-III spectra.

$^b$The measurements from the SDSS-I/II spectra.
\end{table}

\begin{figure*}
 \centering
 \vspace{3ex}
\includegraphics[width=18 cm,height=5 cm]{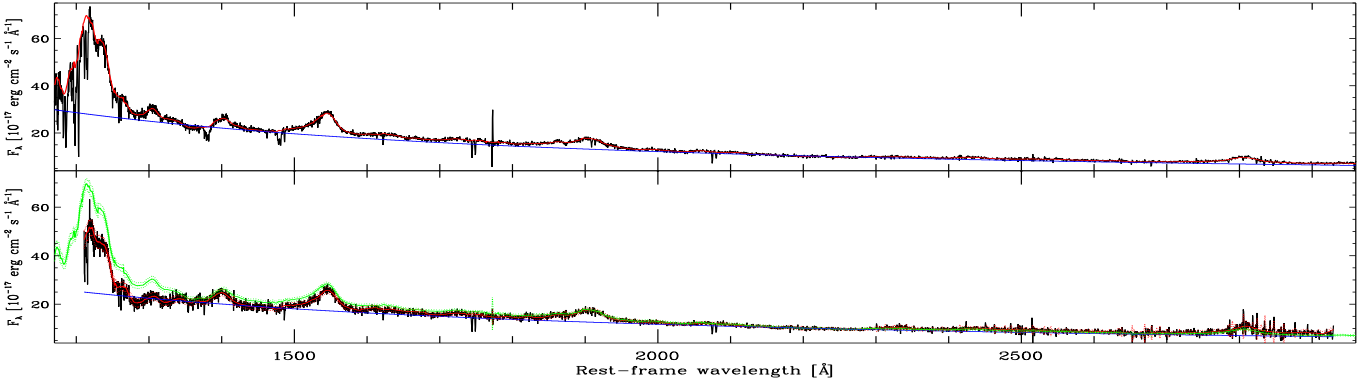}
\vspace{2ex} \caption{The spectra of quasar J095254.10+021932.8 with
$\rm z_e=2.1473$. The blue solid lines represent the power-law
continuum fittings. The lower panel is obtained by SDSS-I/II on 30
December 2000, and the upper panel is obtained by SDSS-III on 25
March 2011. The red solid lines represent the pseudo-continuum
fittings, and the green solid line shown in the lower panel is just
the pseudo-continuum plotted in the upper panel. The green/red
dot-lines represent the corresponding pseudo-continuum fluxes
accounting for the corresponding $\pm 1\sigma$ flux uncertainties.}
\end{figure*}

\begin{figure*}
 \centering
\includegraphics[width=18 cm,height=5.5 cm]{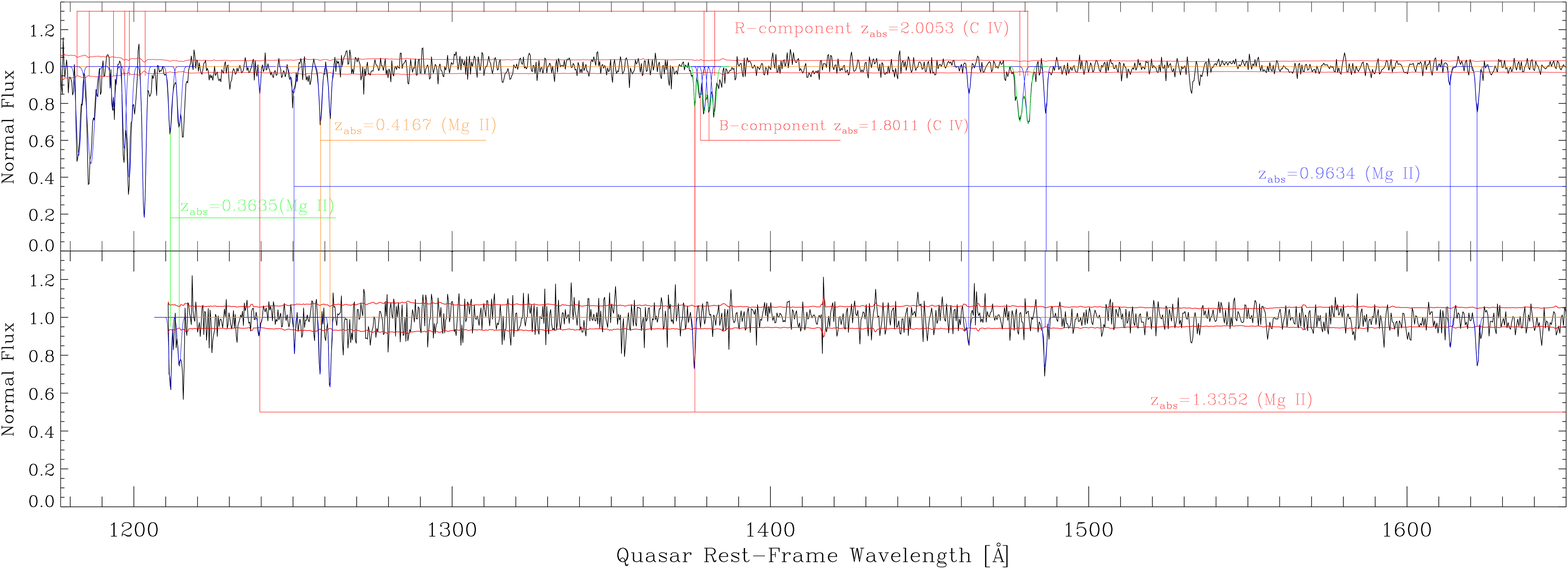}
\vspace{2ex}\vspace{1ex}
\includegraphics[width=18 cm,height=5.5 cm]{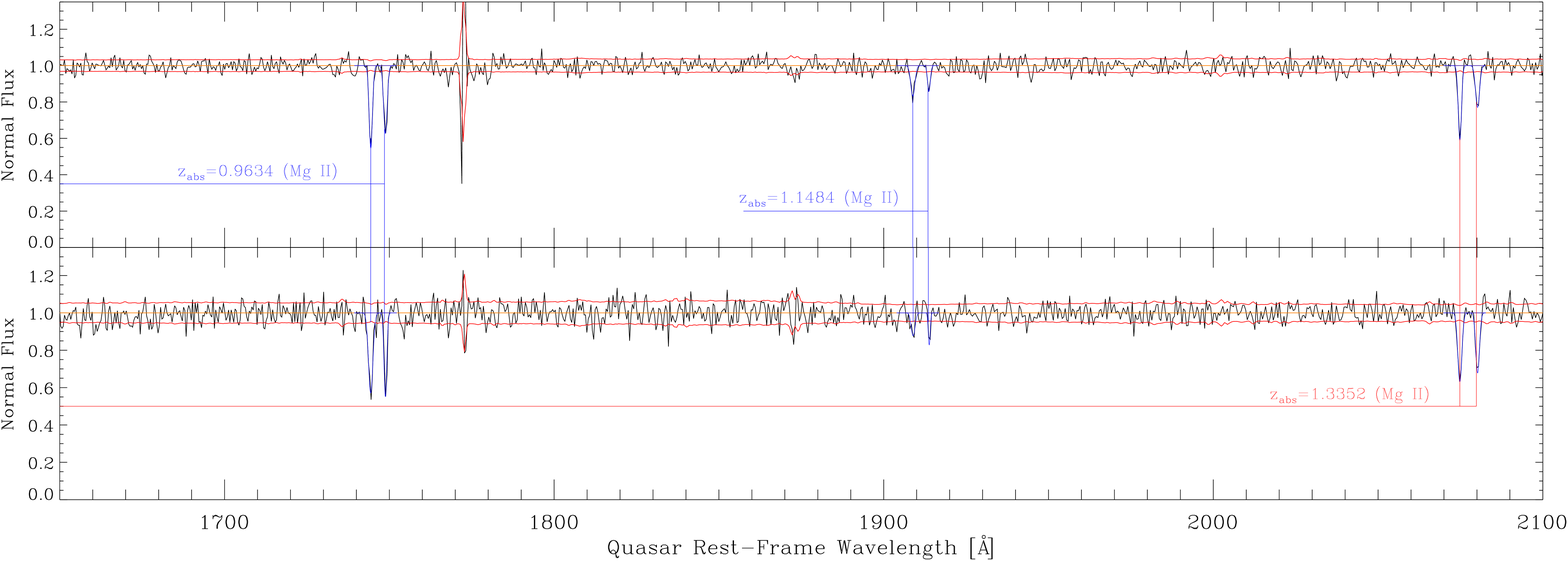}
\vspace{2ex} \caption{The pseudo-continua normalized spectra of
quasar J095254.10+021932.8. The lower panel is obtained by
SDSS-I/II, and the upper panel is obtained by SDSS-III. The red
cures represent the flux uncertainty level which have been
normalized by the corresponding pseudo-continua, the blue curves
represent the Gaussian fittings. The green curves are just the sum
of the multiple lines. The $R-component$ is the C IV absorption
system with $z_{\rm abs}=2.0053$, and The $B-component$ is the C IV
absorption system with $z_{\rm abs}=1.8011$. We also detect 5
$Mg~II~\lambda\lambda2796,2803$ absorption systems with $z_{\rm
abs}=0.3635,~0.4167,~0.9634,~1.1484,~and~1.3352$.}
\end{figure*}

\section{Discussions}
\subsection{Coverage fraction}
As we know from atomic physics, the optical depth ratio of the $\rm
C~IV~\lambda\lambda1548,1551$ doublet is $\tau_{\rm 1548}:\tau_{\rm
1551}\approx2:1$ (Savage \& Sembach 1991; Verner et al. 1994). If
the absorber completely occults the background emission sources, the
optical depth ratio of the resonance doublet should be consistent
with the value expected from the atomic physics. When the absorber
partially occults the background line sources, the unocculted
fluxes, perhaps together with the local emission of the absorber and
the background pho-tons scattered into our sightline, will result in
a deviation of the optical depth ratio from the theoretical value
(e.g., Wampler et al. 1995; Barlow \& Sargent 1997; Hamann et al.
1997). In principle, the intrinsic absorber is often expected to
partially cover the background emission sources, and the coverage
fraction can be calculated from the residual intensities of the
resonance doublet.

The effective coverage fraction ($C_{\rm f}$) accounts for the
fraction of background photons that are obstructed by the absorbing
gas at a given wavelength, and the effective optical depth ($\tau$)
ofthe absorbing gas is considered to be the photons that survive
when the background photons pass through the absorbing gas.
Considering the effective coverage fraction and the effective
optical depth, the normalized residual intensity at a given
wavelength is
\begin{equation}
R(\lambda)=[1-C_{\rm f}(\lambda)]+C_{\rm f}(\lambda)e^{\rm
-\tau(\lambda)}
\end{equation}
For the $\rm C~IV\lambda\lambda1548,1551$ doublet that has an
optical depth ratio of $2:1$, the effective coverage fraction can be
expressed as
\begin{equation}
C_{\rm f}(\lambda)=\frac{[R_{\rm r}(\lambda)-1]^2}{R_{\rm
b}(\lambda)-2R_{\rm r}(\lambda)+1}
\end{equation}
where the subscript $r$ and $b$ refer to the redder and bluer
mem-bers of the $\rm C~IV\lambda\lambda1548,1551$ doublet (see
Hamann et al. 1997; Barlow \& Sargent 1997; Ganguly et al. 1999;
Misawa et al. 2007, for more detail).

It is well known that the emission from active galactic nuclei comes
from multiple, separate components with different structures, such
as the accretion disc, the broad emission-line region (BELR) and the
narrow emission-line region (NELR). It is possible that the
intrinsic absorber shows different coverage fractions for the
different emission sources. For example, the absorber might
partially cover the accretion disc but not cover the NELR, because
the NELR is likely to be far away from the accretion disc. If we
consider that the background photons, which pass through the
absorber, are only from the continuum emission source and the BELR,
and if we assume that the optical depths are the same from the two
emission sources, the effective coverage fraction is the weighted
average of the coverage fractions of the two emission sources, that
is,
\begin{equation}
C_{\rm f}=\frac{C_{\rm c}+WC_{\rm e}}{1+W}
\end{equation}
Here, $C_{\rm c}$ and $C_{\rm e}$ are the coverage fractions of the
continuum emis-sion source and the BELR, respectively, and $W=f_{\rm
e}/f_{\rm c}$ is the ratio of the broad-line flux-subtracted
continuum flux to the continuum flux at the wavelength of the
absorption line (e.g., Ganguly et al. 1999; Misawa et al. 2007; Wu
et al. 2010). For background photons aris-ing from more emission
sources, see the discussions and the more general formalisms by
Gabel et al. (2005) and Misawa et al. (2007). We can compute $C_{\rm
f}$ of the $\rm C~IV\lambda\lambda1845,1551$ absorber using the
equation (2), and we can determine the value of $W$ by measuring the
strength of the broad emission line and the power-law continuum
(which roughly depicts the emission from the accretion disc from UV
to optical) at the position of the doublet. The values of $C_{\rm
c}$ and $C_{\rm e}$ cannot be determined independently of each
other, but we can derive a relation between them using the values of
$C_{\rm f}$ and $W$.

For narrow absorption lines, it is inappropriate to compute the
effective coverage fraction pixel by pixel using the low-resolution
spectra of SDSS. Therefore, here we estimate the values of $C_{\rm
f}$ of the $R-$ and $B-components$ using the normalized residual
intensities at line cores. Thus, we obtain $C_{\rm f}=0.30\pm0.05$
for the $R-component$ and $C_{\rm f}=0.17\pm0.04$ for the
$B-component$.

The continuum emission from the accretion disc from UV to op-tical
can be roughly described with a power-law function
($f\propto\nu^{-\alpha}$). Adopting the method provided by Chen et
al. (2009), we fit the power-law continuum by selecting several
spectral regions without obvious emission lines, as shown in Fig. 1
by blue solid lines. We obtain $\alpha=-1.68$ for the SDSS-III
spectrum and $\alpha=-1.51$ for the SDSS-I/II spectrum. It can be
seen from the upper panel of Fig. 1 that the value of $W$ would be
less than 0.1 for both the $R-$ and $B-components$. Adopting
$W=0.1$, together with the derived values of $C_{\rm f}$, we can
constrain the relations between the $C_{\rm c}$ and $C_{\rm e}$,
which are plotted in Fig. 3.

\begin{figure}
 \centering
\includegraphics[width=7 cm,height=5.5 cm]{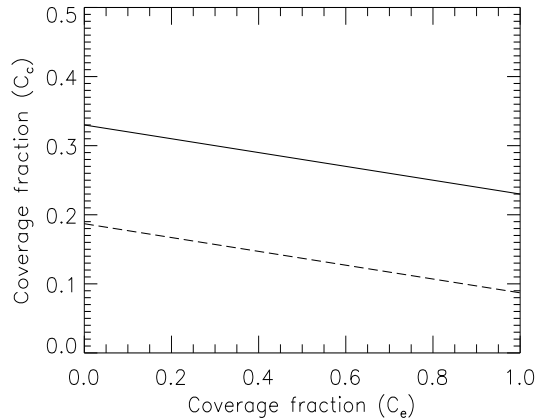}
\vspace{6ex}\caption{The $C_{\rm c}$ --- $C_{\rm e}$ parameter
plane. The solid line is for the $R-component$, and the dash line is
for the $B-component$.}
\end{figure}

\subsection{The properties of the emergence absorption systems}
The time variation of the line strength and the partial coverage are
the two most popular indicators to determine whether the absorber is
truly intrinsic to the quasar. It can be clearly seen from Figs 1
and 2 that both the R-component and B-component show dramatic
variations between the two SDSS observations. These dramatic
variations, together with the effective coverage fractions that have
been derived in Section 3.1, suggest that both the R-component and
the B-component are likely to be intrinsic to the quasar.

The velocity offsets of the $R-component$ ($z_{\rm abs}=2.0053$) and
the $B-component$ ($z_{\rm abs}=1.8011$), with respect to the quasar
($z_{\rm e}=2.1473$), are $\sim 13,800~{\rm km~s^{-1}}$ and $\sim
34,700~{\rm km~s^{-1}}$, respectively. Rough estimations from the
curve of growth give rise to the column densities, $N_{\rm H}$,
would be greater than $10^{\rm 17}~{\rm cm^{-2}}$ for both the $R-$
and $B-components$.

The quasar J095254.10+021932.8 was observed by SDSS-I/II on 2000
December 30, and by SDSS-III on 2011 March 25. The time interval of
the two observations is 1186.4 d at the quasar rest frame. Because
of the difference in focus of the BOSS quasars and stan-dard stars,
with respect to the SDSS-I/II spectra, the BOSS spectra usually
exhibit excess flux at the bluer end and decrescent flux at the
redder end (see fig. 5 of P\^aris et al. 2012). This might explain
the excess flux of the SDSS-III spectrum, with respect to the
SDSS-I/II spectrum, at the blue end of the quasar
J095254.10+021932.8 (see the red and green solid lines in the lower
panel of Fig. 1). Taking into account this discrepancy, the quasar
emissions (the pseudo-continua shown in Fig. 1) are stable for the
two SDSS observations. We have derived the spectral indices of the
power-law continuum for the SDSS-III and SDSS-I/II spectra,
$\alpha=-1.68$ and $\alpha=-1.51$, respectively. We believe that the
discrepancy between the spectral indices can be mainly attributed to
the excess flux at the bluer end and the decrescent flux at the
redder end of the SDSS-III spectrum, rather than to the essence of
the quasar. Therefore, the continuum emissions of the quasar
J095254.10 +021932.8 should be very stable for the two SDSS
observations. Together with the stable pseudo-continua, the
emissions of the broad lines ($\rm Si~IV,~C~IV,~C~III$ and $\rm
Mg~II$) should also be stable at the two epochs. In addition, when
the bolometric luminosity ($L_{\rm bol}$) of the quasar is simply
related to the monochromatic luminosity (e.g. the luminosity at
$\lambda1350$, Chen et al. 2011; Shen et al. 2012), the mass
accretion rate would be also stable (\.M$\propto L_{\rm bol}$).

Time variations of the intrinsic absorption lines seem to be common,
but extreme events of the absorption-line emergence are very rare.
We are aware that there have only been five previous reports that
have observed the C IV absorption-line emergences via optical
spectra (Ma 2002; Hamann et al. 2008; Leighly et al. 2009; Krongold
et al. 2010; Rodr\'Iguez Hidalgo et al. 2011). All these studies
have attributed the emergence events to the absorbers moving across
the quasar sightlines. What are the origins of the R-component and
B-component emerging from the spectrum of the quasar
J095254.10+021932.8?

The question is whether the changes in the ionization state of the
absorbing gas give rise to the emergence of absorption line. These
changes can be caused by the following: (i) an increase in the
incident flux, giving rise to photoionization from the lower
ionization state to the higher ionization state ($\rm C~III
\rightarrow C~IV$); (ii) a decrease in the incident flux, causing
recombination from the higher ionization state to the lower
ionization state ($\rm C~V \rightarrow C~IV$). The intrinsic
absorption lines are often believed to originate in the outflow gas
lifted off the accretion disc, and the $\rm C~IV$ broad emission
line region (Murray et al. 1995; Murray \& Chiang 1997; Proga et al.
2000; Everett 2005). Here, we believe that the emergences of the C
IV absorption doublets are unlikely to be a result of the ionization
changes in the absorbing gas for the following reasons: (i) there is
no evidence of remarkable changes in the observed power-law
continuum; (ii) there is a lack of variation in the broad emission
lines.

The new outflow, arising from the situation where there was
previously no outflow at all, seems to be a reasonable explanation
for the emergence of $\rm C~IV$ absorption lines. Using equation (2)
(i.e. reverberation mapping) of Kaspi et al. (2007), and together
with the monochromatic luminosity at 1350 \AA,~ we can estimate the
radius of the $\rm C~IV$ BELR as $R_{\rm C~IV} \approx 0.33~{\rm
pc}$. Assuming that the $\rm C~IV$ absorption lines form at a radii
just beyond the $\rm C~IV$ BELR, the characteristic flow times ($t
\sim R_{\rm C~IV}/v_{\rm r}$, where the $v_{\rm r}$ is the relative
velocity of the absorber with respect to the quasar) would be
$t_{\rm 1}\sim23.7$ years for the $R-component$, and $t_{\rm
2}\sim9.4$ years for the $B-component$. These flow times are much
larger than the time interval of the two SDSS observations ($\sim
3.3$ years). In addition, the stabili-ties of the power-law
continuum emission and mass accretion rate might imply that there is
no remarkable change in the structure of the accretion disc, which
suggests that there is no new outflow. There-fore, the rise of a new
outflow can be ruled out as an explanation for the extreme events
surrounding the emergence of $\rm C~IV$ absorption line.

The movement of the absorbing gas across the quasar sight-line
probably accounts for the emergence of absorption troughs. The
incomplete occultation of the background emission sources (see
Section 3.1) implies that the sizes of the absorbers are small.
Considering the case of a geometrically thin accretion disc and an
optical thick accretion disc, most of the UV continuum radiation of
the quasar is expected to arise from the inner accretion disc, whose
size is of the order of $D_{\rm cont} \sim 5R_{\rm S}=10GM_{\rm
BH}/c^2$ (Wise et al. 2004; Misawa et al. 2005; Chen et al. 2013).
We directly adopt the virial black hole mass We directly adopt the
virial black hole mass $M_{\rm BH}=10^{\rm 9.56}~M_\odot$, which is
estimated based on the $\rm C~IV$ broad emission line (Shen et al.
2011), as the mass of the black hole. This results in $D_{\rm
cont}=5.35\times 10^{\rm 15}~{\rm cm}$. It can be seen from the
$C_{\rm c}$ --- $C_{\rm e}$ plane (Fig. 3) that the coverage
fraction of the absorber to the continuum emission source is $23\%
\sim 33\%$ for the $R-component$, and $9\% \sim 19\%$ for the
$B-component$.  Thus, we have a characteristic absorber radius of
$\sim 1.5\times10^{\rm 15}~{\rm cm}$ for the $R-component$, and
$\sim 0.7\times10^{\rm 15}~{\rm cm}$ for the $B-component$. Adopting
the time interval of the two SDSS observations (i.e. 1186.4 d) as
the upper limit of the transit time of the absorbing gas, we can
estimate the lower limits of the transverse velocity perpendicular
to the quasar sightline as $\sim 146$ $\rm km~s^{-1}$ and $\sim 73$
$\rm km~s^{-1}$ for the $R-$ and $B-components$, respectively.
Taking the virial black hole mass and assuming that the shift
velocities of the absorbing gas do not exceed the escape velocities,
we can derive the upper limits of the absorber distances relative to
the central region: the absorber with $z_{\rm abs}=2.0053$ locates
at a radius of $r \sim 0.82~{\rm pc}$ and the absorber with $z_{\rm
abs}=1.8011$ at a radius of $r \sim 0.13~{\rm pc}$. From our
estimation, we know that the radius of the
 $\rm C~IV$ BELR is $R_{\rm C~IV} \sim 0.33~{\rm pc}$. In this
way, these absorbers would locate at the vicinity of the BELRE.

Radiation pressure, magnetohydrodynamic and thermal driving are all
mechanisms that can be used to derive an outflow (Proga 2007). The
question is what mechanism dominates the high-velocity outflow of
the quasar J095254.10+021932.8. However, it is beyond the scope of
this paper to accurately derive this mechanism. Ganguly et al.
(2007) and Ganguly and Brotherton (2008) have found an up-per
envelope curve for the relation between the maximum velocity of the
outflow and the quasar luminosity (i.e. $v_{\rm max}\propto L^{\rm
0.662}$), which is expected from the radiation-driven outflow. We
plot the $v_{\rm max}$ --- $L$ plane in Fig. 4 with the upper
envelope curve. It is clear that the max-imum velocity of the
outflow of the quasar J095254.10+021932.8 is close to the upper
envelope curve, which implies that the outflow might be dominated by
radiation pressure.

\begin{figure}
 \centering
\includegraphics[width=7 cm,height=5.5 cm]{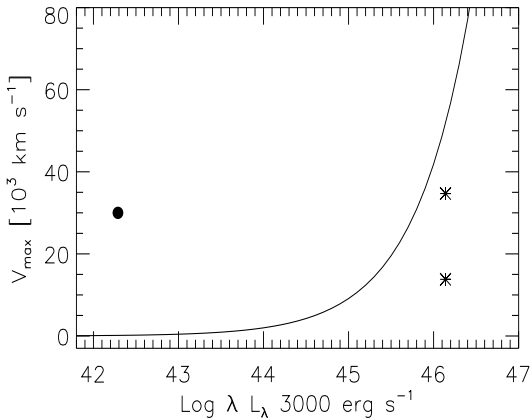}
\vspace{6ex}\caption{The maximum velocity of absorption vs the
monochromatic luminosity at 3000 \AA.~ The stars are for the data of
this work, and the fill circle is for the data in Gupta et al.
2013b.}
\end{figure}

Absorbers that are weakly ionized (e.g., $\rm C~IV$, $\rm N~V$, $\rm
O~VI$)  are often observed in X-ray and UV spectroscopic surveys,
but no highly ionized absorbers (e.g., $\rm Mg~VI$, $\rm Fe~XVII$)
have been detected in UV spectra. The absorption lines with low
ionizations and low velocities, which are observed in both X-ray and
UV spectra and have similar velocities, imply an underlying relation
between the absorption lines in the UV spectra and those in the
X-ray spectra (e.g., Mathur et al. 1994, 1995; Kaspi et al. 2002;
Krongold et al. 2003; Gupta et al. 2013a). These absorption lines
might be formed in a single outflow with multiple discrete
components (e.g. Krongold et al. 2003, 2005, 2007). The component
that is highly ionized gives rise to absorption lines at high
energies, and the component that is weakly ionized produces X-ray
and UV absorption lines. The narrow absorption-line outflows with
ultra-high velocity and high ionization seem to be observed more
often in X-ray spectra (e.g. Tombesi et al. 2010a,b, 2011a,b; Gupta
et al. 2013b). These absorbers would be compact, located close to
the black holes and would have values of column density as large as
$N_{\rm H}={\rm 10^{22}-10^{24}~cm^{-2}}$. We note from fig. 6 of
Gupta et al. (2013b) that most of these ultra-fast outflows lie well
above the upper envelope curve. This suggests that the mechanism
driving these ultra-fast outflows might differ from that of the
out-flows that lie well below the upper envelope curve.

Recently, Gupta et al. (2013b) have become the first to detect an
ultra-fast outflow in the soft X-ray spectra of Ark 564 using the
$\rm O~VI$ absorption line at the velocity of $\sim 0.1c$. This $\rm
O~VI$ absorption line has a similar velocity and ionization level as
the $\rm C~IV$ absorption lines that we have detected in this work.
The value of the column density ($N_{\rm H}$) of this $\rm O~VI$
absorber is $\rm 10^{20}~cm^{-2}$, which is much larger than the low
limit value of the $\rm C~IV$ absorbers ($\rm 10^{17}~cm^{-2}$).
Adopting the bolometric correction factor, one can roughly compute
the monochromatic luminosity at 3000 \AA~ from the bolometric
luminosity of the $Ark~564$ (Shen et al. 2011), and plot the $\rm
O~VI$ ultra-fast outflow in the $v_{\rm max}$ --- $L$ plane (see the
filled circle shown in Fig. 4). We can see from Fig. 4 that the $\rm
O~VI$ ultra-fast outflow lies well above the upper envelope curve,
which is significantly different from the $\rm C~IV$ outflow of this
work. Therefore, the magnetohydrodynamic mechanism might dominate
the $\rm O~VI$ ultra fast outflow (Gupta et al. 2013b), unlike the
radiation-pressure mechanism linked to the $\rm C~IV$ outflow.
Although some properties of the O VI ultra-fast outflow and the C IV
outflow display obvious differences, the similar velocities and
ionization levels might lead us to investigate the connection
between the absorption lines in the UV spectra and those in the
X-ray spectra. This precise connection would be complicated and is
beyond the scope of this paper.

\section{Conclusions}
The quasar J095254.10+021932.8 with $z_{\rm e}=2.1473$ was observed
by SDSS-I/II on 30 December 2000, and reobserved by SDSS-III on 25
March 2011. The time interval of the two observations is 1186.4 days
at quasar rest-frame. We find that two $\rm
C~IV\lambda\lambda1548,1551$ narrow absorption systems, which are
respectively located at $z_{\rm abs}=2.0053$ ($R-component$) and
$z_{\rm abs}=1.8011$ ($B-component$), are obviously imprinted on the
SDSS-III spectrum. However, these two absorption systems can not be
observed from the SDSS-I/II spectrum. With respect to the quasar,
the velocities offset is $\rm \sim 13,800~km~s^{-1}$ for the
$R-component$, and $\rm \sim 34,700~km~s^{-1}$ for the
$B-component$. We also find that both the two absorbers partially
occult the background emission sources. The effective coverage
fraction is $C_{\rm f}=0.30\pm0.05$ and $C_{\rm f}=0.17\pm0.04$ for
the $R-component$ and $B-component$, respectively.

The stabilities of the continuum radiations and the broad emission
lines of the quasar J095254.10+021932.8 suggest that the dramatic
variations of absorption troughs are unlikely to be caused by the
changes in the ionization state of the absorbing gas. In addition,
considering the time interval of the two SDSS observations, a new
outflow scenario, arising from the situation where there was
pre-viously no outflow at all, cannot account for the line
variations of both the R-component and B-component.

The multiple streaming gas moving across our sightline might be the
best explanation for the emergence of the two C IV absorption
systems. Adopting the time interval of the two SDSS observations as
the upper limit of the absorber transit time, we estimate the lower
limits of the transverse velocity perpendicular to the line of
sight. This gives rise to velocities of $\rm 146~km~s^{-1}$ for the
$R-component$ and $\rm 73~km~s^{-1}$ for the $B-component$. Assuming
that the shift velocities of the absorbers do not exceed the escape
velocities, the distances of the absorbers relative to the central
region are $\rm \sim 0.82~pc$ for the $R-component$, and $\rm \sim
0.13~pc$ for the $B-component$. These absorbers might lie in the
vicinity of the broad emission line region of the quasar ($\rm
R_{C~IV} \sim 0.33~pc$).

\vspace{6mm}We thank the anonymous referee for helpful comments and
suggestions. This work was supported by the Guangxi Natural Science
Foundation (2012jjAA10090), the National Natural Science Foundation
of China (No. 11073007), and the Guangzhou technological project
(No. 11C62010685).

\end{document}